\documentclass[pre,twocolumn,showpacs,showkeys]{revtex4}

\usepackage{graphicx}
\usepackage{dcolumn}
\usepackage{bm}

\usepackage{epsfig}
\usepackage{amssymb}

\newcommand{\D}{\mathrm{d}}
\newcommand{\E}{\mbox{1\hspace{-3 pt}I}}
\newcommand{\e}{\vec{\text{e}}}

\newcommand{\R}{\mathbb{R}}
\newcommand{\Exp}[1]{\text{e}^{#1}}
\newcommand{\binomial}[2]{\left(\begin{array}{c} #1 \\ #2 \end{array} \right)}
\newcommand{\pd}{\partial}
\begin{document}
\title{Chaoticity of the Wet Granular Gas}

\author{A. Fingerle}
\email{axel.fingerle@ds.mpg.de}
\author{S. Herminghaus}
\email{stephan.herminghaus@ds.mpg.de}
\author{V. Yu. Zaburdaev}
\email{vasily.zaburdaev@ds.mpg.de}
\affiliation{Max Planck Institute for Dynamics and Self-Organization\\
Bunsenstr. 10, Germany - 37073 Goettingen}

\date{\today}

\begin{abstract}
In this work we derive an analytic expression for the
Kolmogorov-Sinai entropy of dilute wet granular matter, valid for
any spatial dimension. The grains are modelled as hard spheres and
the influence of the wetting liquid is described according to the
Capillary Model, in which dissipation is due to the hysteretic
cohesion force of capillary bridges. The Kolmogorov-Sinai entropy
is expanded in a series with respect to density. We find a rapid
increase of the leading term when liquid is added. This
demonstrates the sensitivity of the granular dynamics to humidity,
and shows that the liquid significantly increases the chaoticity
of the granular gas.

\end{abstract}
 \maketitle

\section{Introduction} \label{SecIntro}
The field of granular physics has undergone considerable progress
in recent times \cite{Review2, Poeschel}. As part of soft matter
physics, granulates have inspired the development of
non-equilibrium statistical mechanics \cite{Dufty, Annette}.
Its potential to the foundation of physics can hardly be over
estimated, since granular gases provide a road away from the
well-developed Boltzmann-Enskog theory of conservative gases
towards dissipative systems far from thermal equilibrium. In
connection with geophysics, some aspects of landslides may be
understood in terms of a solid-liquid phase transitions of wet
granular matter \cite{Mario, Schulzes, Stephan}, and wet granular
gases are of technological relevance in granulators, pelletizers,
and other instances in process engineering.

Wet granular gases are systems consisting of mesoscopic particles
and a liquid phase wetting the particles. Despite their
importance, the theory of wet granular matter is still nascent.
There is a growing number of experimental \cite{TaibiLenoblePozo}
and numerical work \cite{Youssoufi} on this subject, but
the hysteretic nature of the liquid bridge interaction was not
taken into account in the modelling. We stress that the attraction
force mediated by capillary bridges is not a function of distance
but depends on the collision history. The theory of wet granular
matter advanced with recent simulation and models describing the
free cooling state \cite{ZBH,FH}. To the best of our knowledge,
the hysteretic dissipative dynamics of wet granular matter was
treated analytically first in \cite{FHZ}. In this article we
elaborate on this approach which treats the wet granulate as a
complex dynamical system and uses powerful tools available in this
area. Such is the Lyapunov spectrum,
\begin{equation}
\lambda_j
=\lim_{t\rightarrow \infty}\frac{1}{t}
\ln 
\frac{\delta\Gamma_j(t)}{\delta\Gamma_j(0)}
\ .
\end{equation}
It gives the rate of exponential divergence or convergence of two
equal copies of the system in phase space, $\delta
\Gamma_j(t)=\Gamma_j^{(1)}(t)-\Gamma_j^{(2)}(t)$, with perturbed
initial conditions $\delta \Gamma_j(0)$. A positive Lyapunov
exponent indicates chaotic behavior, i.e. sensitive dependence on
the initial conditions \cite{Kantz}.
Since we are dealing with a closed system the sum of all positive
Lyapunov exponents equals the Kolmogorov-Sinai entropy (KSE)
\cite{Pes77, Review1}.

The KSE is an indispensable tool in the modern description of
dynamical systems. Firstly, from it we learn about the degree of
chaoticity because its inverse is the time scale of
predictability. Secondly, this dynamical entropy is a well-defined
quantity for both equilibrium and non-equilibrium systems.
Thirdly, when tiny deviations of initial conditions that were not
observable in the beginning are enlarged by the evolution, this
can be interpreted as the production of information about the
initial conditions. Finally, the KSE is known to be related to
macroscopic properties such as transport coefficients
\cite{TC1,TC2a,TC2b,TC2c,TC2d,TC2e,TC3}.

Our objective is to compute the KSE for the wet granular gas.
Pioneering work has been done by H. van Beijeren, J. R. Dorfman
{\it et al.} \cite{Dorfman,Dorfman2} in the analytic treatment of
sums of Lyapunov exponents for the gas of hard elastic spheres. We
develop a generalization of the method suggested in
\cite{Dorfman}.

This article is organized as follows. In section \ref{SecConc} we
describe in detail the hysteretic interaction of wet granulates.
This Capillary Model allows the sticking of particles by
attractive forces in contrast to the ``Standard Model'' for dry
granulates which assumes that a certain fraction of energy is lost
instantaneously by inelastic collisions. In section \ref{SecNPart}
we use the terminology developed in section \ref{SecConc} to
relate the behavior of the two-particle system to the full
$N$-particle system. Thereby we are lead to determine the
probability distribution for colliding pairs of particles in
section \ref{SecAver}. In section \ref{SecExpVelo} we derive the
formula that expresses the expansion of velocity space as a
function of the two-particle initial conditions for arbitrary
spatial dimension. In
section \ref{SecRes} the results of the sections
\ref{SecNPart}-\ref{SecExpVelo} are combined to accomplish the
computation of the KSE.
\section{The Capillary Model} \label{SecConc}
There is an experimentally well confirmed Capillary Model for the
dynamics of wet granulates, that will be applied here
\cite{Mario}. The system consists of hard spherical grains with
equal diameter $\sigma$ and equal mass $m$. These are covered by a
liquid film, so that every time two particles touch, a liquid
bridge is formed. The Capillary Model assumes that bridges are
formed instantaneously. As we focus on the dilute gas, we may
restrict our considerations to pair interactions.

Experiments and computations \cite{Force1, Willett}
yield a capillary force law that is excellently described by
\begin{eqnarray}
F = \frac{\pi \gamma \sigma \cos \theta_{\text{w}}} {1 + 0.74 \
{\mathsf s} + 1.25 \; {\mathsf s}^2} \label{ForceLaw}
\end{eqnarray}
with the wetting angle $\theta_{\text{w}}$, the surface tension
$\gamma$, and ${\mathsf s}=s \sqrt{\sigma/V_{\text{bridge}}}$
being the surface separation $s$ expressed in the natural length
unit $\sqrt{V_{\text{bridge}}/\sigma}$ of the liquid bridge volume
$V_{\text{bridge}}$.
\begin{figure} \begin{center}
\scalebox{0.34}{\epsffile{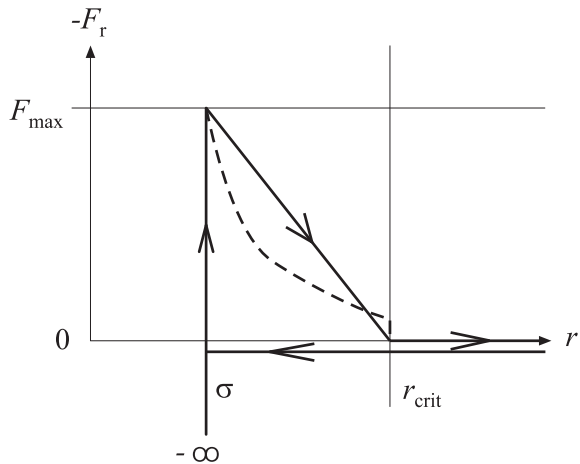}} \caption{Radial forces
between a pair of wetted spheres. Solid line: The radial force of
the Extended Capillary Model is plotted versus the center distance
$r$. There is no interaction between the particles as they
approach. After the collision applies
$\vec{F}(r)=-F_{\text{max}}\frac{r_{\text{crit}}-r}{r_{\text{crit}}-\sigma}
\frac{\vec{r}}{r}$ for $r \in (\sigma,r_{\text{crit}})$, otherwise
there is no force. Dashed line: Experiments yield a decreasing
force law \cite{Force1,Willett} with a discontinuity at the
rupture. Therefore the even simpler Minimal Capillary Model which
assumes a constant force that drops to zero at the critical
separation is a good alternative approximation. The hysteretic
interaction is the relevant property which is described by both
the Minimal and the Extended Capillary Model.}
\label{ExpForce_Fig} \label{LinForce_Fig}
\end{center} \end{figure}
The Capillary Model assumes that the bridge pinches off at a
critical surface separation $s=s_{\text{crit}}$ (i.e. at a
distance $r_{\text{crit}}=\sigma+s_{\text{crit}}$ of the centers).
To leading order, the rupture distance $s_{\text{crit}}$ equals
the cubic root of the bridge volume $V_{\text{bridge}}$. The
energy that was stored in the stretched liquid bridge before the
rupture is dissipated into the liquid and lost for the granular
motion. We emphasize that this is the only dissipative mechanism
in the Capillary Model (cf. the review article \cite{Stephan},
especially Fig.~7 therein, for the capillary regime in which the
Capillary Model applies.) In the moment of the rupture, the system
is non-Hamiltonian because the atomic degrees of freedom of the
liquid to which energy flows are masked out in the description of
the granular dynamics. Of course the forces acting on the grains
are finite at the rupture, so that the trajectories (as functions
of time) are continuous in the granular phase space and
differentiable with respect to the initial state before the
rupture.

By a collision we denote the moment when two particles in the
entire $N$-particle system touch each other. Since we are
interested in statistical statements and a point in time is of
measure zero, we can assume without loss of generality that there
is a unique sequence of collisions. For a certain pair of
colliding particles, we refer to the ``collision cycle'' as the
time interval $[t_{\text{i}},t_{\text{f}}]$ that comprises the
collision of these two particles. The collision cycle starts at
$t_{\text{i}}$ when the last particle of the two breaks free from
its former collision partner and ends at $t_{\text{f}}$ in the
moment when the liquid bridge between them ruptures.

During its collision cycle the radial motion of the two-particle
system traverses a hysteresis loop. This is shown in
Fig.~\ref{LinForce_Fig} for the force (\ref{ForceLaw}) (dashed
line) and for a simpler force law (solid line). The solid line in
Fig.~\ref{LinForce_Fig} falls off linearly with the surface
separation $s$. This is the Extended Capillary Model in contrast
to the Minimal Capillary Model of \cite{Mario} which assumes a
constant force. The corresponding hysteretic ``potential'' of the
Extended Capillary Model is
\begin{eqnarray} \label{potential}
 \frac{\phi(r)}{E_{\text{loss}}} =  \left\{
\begin{array}{ll}
-1, & \sigma < r\ \text{ before first collision,} \\
-\left(\frac{r_{\text{crit}}-r}{r_{\text{crit}}-\sigma}\right)^2,
& \sigma < r \leq r_{\text{crit}} \text{ after collision,} \\
0, & r_{\text{crit}} \leq r \text{ after collision, } \\
\infty, & r < \sigma .
\end{array}
\right.
\end{eqnarray}
In both, the Minimal and the Extended Capillary Model, the
hysteretic loss of energy, i.e. the area
$E_{\text{loss}}=-\int_{\sigma}^{\sigma+s_{\text{crit}}}
F_{\text{r}} \ \D r$ in Fig.~\ref{ExpForce_Fig}, is a
characteristic system property. When the energy in the center of
mass system is below $E_{\text{loss}}$, colliding particles will
form a stable bound state with periodic collisions. With faster
relative motion the liquid bridge exists for a finite time until
the particles scatter off each other. We define a corresponding
relative velocity $v_{\text{loss}}$ by $E_{\text{loss}}={m}
v_{\text{loss}}^2/{4}$ (with the additional factor ${1}/{2}$
because ${m}/{2}$ is the reduced mass). From this point on we
distinguish between scattering events and collisions leading to
bound states. For the scattering, the restitution coefficient
$\epsilon=E_{\text{f}}/E_{\text{i}}$ of the Capillary Model is an
increasing function of the initial energy or velocity:
\begin{eqnarray}
\epsilon(E_{\text{i}})=\sqrt{1-\frac{E_{\text{loss}}}{E_{\text{i}}}}
\text{ or }
\epsilon(v_{\text{i}})=\sqrt{1-\frac{v^2_{\text{loss}}}{v^2_{\text{i}}}}
\ . \label{ResKoeff}
\end{eqnarray}
The binding threshold $E_{\text{loss}}$ of the Capillary Model
contrasts sharply with the widespread models for dry granules that
assume either a constant or with increasing velocity decreasing
coefficient of restitution for the collision of viscoelastic
particles \cite{Poeschel}, \footnote{In the capillary regime
described in \cite{Stephan}, the dissipation by inelastic
collisions \cite{Poeschel} is dominated by the hysteretic liquid
bridge interaction.}.

Let us denote by $v_{\text{crit}}$ the critical modulus of the
relative velocity $\vec{v}_{\text{i}}\equiv\vec{v}_1-\vec{v}_2$,
that determines wether the incoming particles will form a bound
state or scatter. For head-on collisions (impact parameter $b=0$)
$v_{\text{crit}}=v_{\text{loss}}$, otherwise
$v_{\text{crit}}>v_{\text{loss}}$ since there is additional energy
in the rotary motion. The next step is to determine
$v_{\text{crit}}$ as a function of $b$.
\subsection*{Determination of the Critical Velocity}
\begin{figure} \begin{center}
\scalebox{0.34}{\epsffile{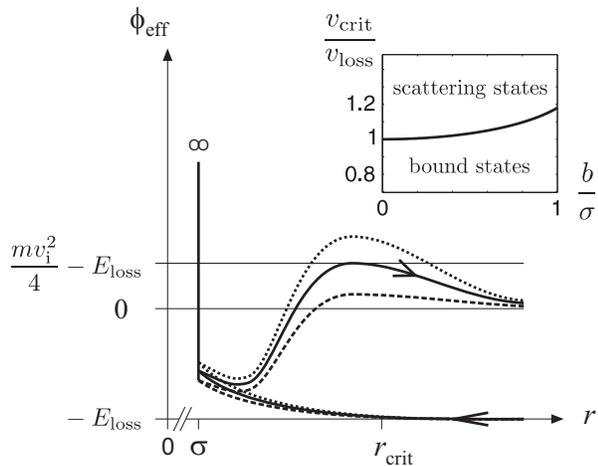}} \caption{The effective
potential for $v_{\text{i}}>v_{\text{loss}}$ and three different
impact parameters. For the solid line in the middle $b$ and
$v_{\text{i}}$ are critical. For the higher $b$ (fine dotted line)
the particles are bound, for a lower $b$ (roughly dotted line)
they scatter. The inset shows the complete space of collision
parameters. The critical velocity $v_{\text{crit}}$ (plotted in
units of $v_{\text{loss}}$ for ${r_{\text{crit}}}=2{\sigma}$) as a
function of the scaled impact parameter ${b}/{\sigma}$ divides the
plane in bound and scattering states.} \label{Gross}
\end{center} \end{figure}

The bridge interaction is a central force problem.
If $v_{\text{i}}$ is lower than $v_{\text{loss}}$, the effective
potential
\begin{equation}
\phi_{\text{eff}}(r)=\frac{mb^2v_{\text{i}}^2}{4r^2}+\phi(r)
\label{EffPot}
\end{equation}
(of the liquid bridge potential given by (\ref{potential})) does
not reach a maximum in $r$ after the collision and leads to a
bound state. For most $v_{\text{i}}>v_{\text{loss}}$ the particles
scatter, but there are some bound cases with high angular momenta,
corresponding to high impact parameters. Figure~\ref{Gross} shows
three effective potentials for a given initial velocity
$v_{\text{i}}$ and different impact parameters $b$. In the case
drawn with solid lines, $b$ and $v_{\text{i}}$ fulfill the
critical relation $v_{\text{i}}=v_{\text{crit}}(b)$. For the
higher $b$ (fine dotted line in Fig.~\ref{Gross}) we have
$v_{\text{i}}<v_{\text{crit}}(b)$ so that a bound system is
formed.
Hence the criterion is that $\phi_{\text{eff}}(r)$ touches the
asymptotic energy $E_{\text{loss}}-{m \, v_{\text{i}}^2}/{4}$ in a
single point. For the Extended Capillary Model it is possible to
calculate these intersections explicitly. These are the roots of
$\left(E_{\text{loss}}-{m \, v_{\text{i}}^2} / {4} +
\phi_{\text{eff}}\right)r^2$, which is a fourth order polynomial
in $r$ with one trivial root at $r=0$ and another unphysical root
for $r<\sigma$. So there are two real roots for the bound state
which turn into a complex conjugated pair of roots for the
scattering state. (Since the derivative of $\phi_{\text{eff}}$ is
continuous and negative at $r=r_{\text{crit}}$, the turning point
$r_{\text{max}}$ of a bound state follows correctly from this
analytic consideration to be $r_{\text{max}}<r_{\text{crit}}$
without the need to take the non-analytic point
$r=r_{\text{crit}}$ of $\phi_{\text{eff}}$ into account.) The
easiest way is to compute the discriminant of the fourth order
polynomial $\left(E_{\text{loss}}-{m \, v_{\text{i}}^2} / {4} +
\phi_{\text{eff}}\right)r^2$, which is equal to
\begin{eqnarray*}
 && 16  v^4 \ \underline{b^4} 
\\
& + &    \left( 8 v^6 - 4 v^4 \left(5 \gamma + 9 \right) +
          v^2 \left(27 + 18 \gamma -
          \gamma^2\right)\right) \ \underline{b^2} 
          \\
& -&  v^6 + v^8
 + 3 v^6 \gamma +
    3 v^4 \left(\gamma-1\right)  \gamma +
    v^2 \left(\gamma - 3 \right) \gamma^2 - \gamma^3 \ ,
\end{eqnarray*}
with
$\gamma=\sigma\frac{2r_{\text{crit}}-\sigma}{\left(r_{\text{crit}}-\sigma\right)^2}$.
The discriminant vanishes as the two physical roots coincide.
Since the impact parameter $b$ enters the problem only trough the
angular momentum term in (\ref{EffPot}), the discriminant is a
quadratic function of $b^2$. Therefore it is elementary to give
$b_{\text{crit}}(v_{\text{i}})$ as the inverse function of
$v_{\text{crit}}(b)$ explicitly:
\begin{widetext}
\begin{equation}
\frac{b_{\text{crit}}(v_{\text{i}})}{\sigma}= \frac{\sqrt{-8 - 20
\delta^2 + \delta^4 + 16 w^2 + 20 \delta^2 w^2 - 8 w^4 - \delta
\left(8 + \delta^2 - 8 w^2\right)^{3/2}}}{4 \sqrt{2} w (\delta-1)}
\ , \label{ECMkrit}
\end{equation}
\end{widetext}
with $\delta=\frac{r_{\text{crit}}}{r_{\text{crit}}-\sigma}$ and
$w=\frac{v_{\text{i}}}{v_{\text{loss}}}$. This function is plotted
as inset in Fig.~\ref{Gross}. Much more concise is the
corresponding function for the Minimal Capillary Model:
\begin{equation}
{v_{\text{crit}}(b)}  = \frac{v_{\text{loss}}}
{\sqrt{1-\frac{b^2}{r^2_{\text{crit}}}}} \ . \label{MCMkrit}
\end{equation}
In the following sections including the main results
(\ref{Expansion})-(\ref{GeneralResult}) of this article, we shall
be completely general without the need to specify for the Minimal
or Extended Capillary Model.
\section{How to Relate the Two-Particle System to the $N$-Particle
System} \label{SecNPart} In the previous section we have shown how
on the level of two-particle interactions the most important
property of the real wet granular gas, namely the hysteretic
binding and breaking of liquid bridges, can be modelled. Further,
we have seen that the bond energy of the liquid bridge gives rise
to the sticking of particles. In this section we treat the
many-particle system.

\begin{figure} \begin{center}
\scalebox{0.55}{\epsffile{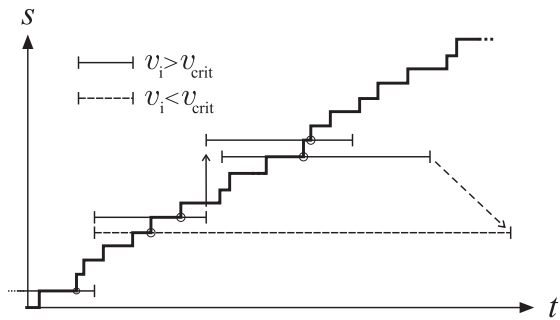}} \caption{The collision
sequence $s(t)$ and the collision cycles: the step function $s(t)$
is the total number of collisions in the entire $N$-particle
system until time $t$. The horizontal solid and dashed bars
symbolize the collision cycles for scattering and bound pairs
respectively. For the derivation is important that overlapping
cycles affect different pairs of particles. The dashed arrow
indicates a third particle that hits and breaks up a bound
two-particle state. } \label{TimeLine}
\end{center} \end{figure}

Let $\nu$ denote the mean collision frequency per particle. If the
modulus of the initial relative velocity $v_{\text{i}}$ is lower
than $v_{\text{crit}}$, so that particles stick together, the
collision cycle is not terminated until a third particle bumps
into the bound two-particle system. We assume that the outstate of
such a three-particle event contains free particles, because the
formation of higher mass clusters is rare in the gas-like state
(cf. Fig.~\ref{CoordinationPlot}).
The pair interactions taking place in the $N$-particle system may
be envisaged as shown in Fig.~\ref{TimeLine}.
The number of collisions up to time $t$ is denoted by $s(t)$.
Since $s(t)$ is strictly monotonic its inverse $t(s)$ exists. The
collision rate of the system, ${s}/{t(s)}$, tends for $s
\rightarrow \infty$ to ${N \nu}/{2}$ (each collision involves two
particles). To have the steps visible Fig.~\ref{TimeLine} has been
drawn for low $N$. The horizontal bars represent the concept of
collision cycles introduced in the last section. There are two
particles which are going to collide. As the beginning of the
collision cycle we take the time when the last of these two
particles has ruptured its liquid bridge connection to some
previous collision partner. The collision cycle will end when
these two particles rupture the liquid bridge between them. Thus a
solid arrow in Fig.~\ref{TimeLine} shows that one of the particles
which just finished its collision cycle immediately begins another
one. The dashed arrow indicates that a third particle (that came
out of another collision cycle) ends a bound two-particle state.

With this picture in mind the computation of the KSE can be
tackled. As stated by Pesin's theorem the KSE equals the sum of
all positive Lyapunov exponents, because the system is closed and
sufficient chaotic \cite{Pes77}. Lyapunov exponents describe the
rate at which a certain direction in phase space grows or shrinks
for large times. There is a orthogonal set of Lyapunov vectors
$\xi_j$ describing the direction while the associated Lyapunov
exponent $\lambda_j$ describes the exponential rate
\begin{equation}
\xi_j(t) \simeq  \xi_j(0) \ \Exp{\lambda_j \; t}
\end{equation}
for long times $t$. According to the sign of $\lambda_j$ one
speaks of stable or unstable directions. The deviations in the
initial conditions are infinitesimal small, i.e. the Lyapunov
exponents characterize the tangent space map associated with a
certain trajectory. In an ergodic system the Lyapunov spectrum
$\{\lambda_j\}$ is independent of the trajectory according to
Oseledec's theorem \cite{Oseledec}. There is no doubt about the
ergodicity of the gas of $N\gg 1$ hard spheres \cite{Simanyi}.

Since in a dilute system the free flight time and the mean free
path are large compared to the interaction time and the range
$r_{\text{crit}}$ of the interaction, perturbations of velocities
are amplified as compared to spatial deviations \cite{Dorfman}.
This is not to be understood as a neglect of the spatial Lyapunov
exponents. The Capillary Model is symplectic \cite{FH} so that for
each positive exponent $\lambda_j$ there is a negative exponent
$\lambda_k=-\lambda_j$ and the fact that the spatial deviations
remain small means that the spatial directions
mainly contain negative Lyapunov exponents, while the positive
ones are assigned to velocities. So the conjecture is that the
velocity space coincides (approximately) with the unstable
manifold of the system. Based on this conjecture the KSE,
$h_{\text{KS}}$, is given by the logarithmic volume growth rate in
velocity space:
\begin{equation}
h_{\text{KS}}= \lim_{s \rightarrow \infty} \frac{1}{t(s)} \
\ln\left\vert \det \prod_{i=1}^s M_i \right\vert \ . \label{KSe1}
\end{equation}
The deviation matrix $M_i$ of the $i$'s collision cycle is
restricted to velocity space, so that it describes the evolution
of velocity perturbations. There are three crucial points here:
(i) This limit exists by virtue of Oseledec's multiplicative
ergodic theorem \cite{Oseledec}.
(ii) We have an unique collision sequence.
(iii) Although there are pair interactions occurring with time
overlaps, there is no ordering problem when writing down the total
deviations as a product of collision cycles, because the
coexisting liquid bridge interactions affect always disjoint pairs
(by the assumption that there are two-particle clusters only) and
deviation matrices of disjoint pairs commute. Therefore
the matrices $M_i$ can 
describe the full collision cycle of a single pair of particles,
ignoring all other interactions taking place simultaneously in the
$N$-particle system.
Our approach differs from \cite{Dorfman}, because the Capillary
Model has a hysteretic interaction with finite interaction time.
The dry limit follows by turning off the interaction,
$E_{\text{loss}}\rightarrow 0$, as a special case.


The expression (\ref{KSe1}) can be simplified dramatically:
\begin{eqnarray}
\frac{h_{\text{KS}}}{N}&=& \frac{1}{N} \lim_{s \rightarrow \infty}
\frac{1}{t(s)} \ \ln\left\vert \det \prod_{i=1}^s M_i \right\vert \nonumber \\
&=&\frac{1}{N} \lim_{s \rightarrow \infty} \frac{1}{t(s)} \
\sum_{i=1}^s \ln \left\vert \det  M_i \right\vert  \nonumber  \\
&=&\frac{1}{N} \lim_{s \rightarrow \infty} \frac{s}{t(s)} \
\frac{\sum_{i=1}^s\ln\left\vert \det M_i \right\vert}{s}  \nonumber \\
&=&\frac{\nu}{2} \ \left<  \ln \left\vert \det M \right\vert
\right> \ .
 \label{KSe2}
\end{eqnarray}
Herein the brackets $<\dots>$ denote averaging over the
two-particle phase space only.

Since we expect the Lyapunov exponents to be of the order of the
collision frequency $\nu$, they are (according to the limit in
(\ref{KSe1})) only well-defined if we let the system evolve for a
time
\begin{eqnarray*}
  t_{\text{Lyapunov}} \gg \frac{1}{\nu} = t_{\text{coll}}\ .
\end{eqnarray*}
In the subsequent discussion we will point out that this can be
fulfilled even if there was no external driving mechanism to keep
the dissipative system in a stationary state. Clearly, without a
thermostat the system cools, $\dot T < 0$, \cite{ZBH, FH}. The
collision frequency $\nu$ is of the order ${\vert \dot T
  \vert}/{E_{\text{loss}}}$.
On the other hand, cooling will be irrelevant on time scales below
$ t_{\text{cool}}={T} / {\vert \dot T \vert}$ . So the hierarchy
\begin{eqnarray*}
t_{\text{coll}} \ll  t_{\text{Lyapunov}} \ll t_{\text{cool}}
\end{eqnarray*}
of time scales can be fulfilled if
\begin{equation}
E_{\text{loss}} 
\ll T \ . \label{OhneThermo}
\end{equation}
This implies that for weak liquid bridges as compared to the
thermal energy we may speak of a Lyapunov spectrum independently
from the question of the thermostat. No additional limitation is
set, since the condition (\ref{OhneThermo}) is already required to
be consistent with the gas state (displaying mainly single
particles instead of clusters) which is studied in this work.

Two tasks remain. The determination of the probability
distribution for the formula (\ref{KSe2}) is done in the next
section. To make use of momentum conservation the subspace is
spanned by the center of mass position
$\vec{R}\equiv\frac{\vec{r}_1+\vec{r}_2}{2}$ and velocity
$\vec{V}\equiv\frac{\vec{v}_1+\vec{v}_2}{2}$ of the two-particle
system, as well as the distance $\vec{r}\equiv\vec{r}_1-\vec{r}_2$
between the centers of the spheres and their relative velocity
$\vec{v}\equiv\vec{v}_1-\vec{v}_2$. The last step is to compute
for any spatial dimension $D$ the matrix $M$ appearing in
(\ref{KSe2}), which maps for a specific point in the
$4D$-dimensional phase space $(\vec{R},\vec{r},\vec{V},\vec{v})$
the initial velocity deviations
\begin{equation}
\binomial{\delta \vec{V}_{\text{i}}}{\delta \vec{v}_{\text{i}}}
\nonumber
\end{equation}
from the beginning of the collision cycle to the final deviations
\begin{equation} \binomial{\delta
\vec{V}_{\text{f}}}{\delta \vec{v}_{\text{f}}} = M
\binomial{\delta \vec{V}_{\text{i}}}{\delta \vec{v}_{\text{i}}}
\label{DefM}
\end{equation} at the end of the collision cycle. This is done in
section \ref{SecExpVelo}.

Before we derive the joint probability density a comment on the
velocity distribution itself is in order. It is well-known that
for dissipative gases the velocity distribution can deviate from
the Maxwell-Boltzmann velocity distribution \cite{Herbst}
depending on the state and driving mechanism.
For explicit results we shall use the Maxwell-Boltzmann velocity
distribution,
\begin{eqnarray}
P(v_1,v_2) \ \D^D v_1 \ \D^D v_2 &=&
\left(\frac{\alpha}{\pi}\right)^{D} \ \Exp{-\alpha(v_1^2+v_2^2)}
                    \ \D^D v_1 \ \D^D v_2 \nonumber \\
&=& \left(\frac{\alpha}{\pi}\right)^{D} \
\Exp{-\alpha(2V_{\text{i}}^2+\frac{1}{2}v_{\text{i}}^2)}
                    \ \D^D V_{\text{i}} \ \D^D v_{\text{i}} \nonumber \\
                    &=& P(V_{\text{i}},v_{\text{i}}) \ \D^D V_{\text{i}} \ \D^D v_{\text{i}}
                    \label{VeloSep}
\end{eqnarray}
with $\alpha=\frac{m}{2T}$. The result for the KSE will also be
given in a form that is readily evaluated for any velocity
distribution.
For the distribution (\ref{VeloSep}) the modulus $v_{\text{i}}$ of
the initial relative velocity is distributed according to
\begin{equation}
P(v_{\text{i}}) \ \D v_{\text{i}} =
\frac{2\left(\frac{\alpha}{2}\right)^{\frac{D}{2}}}{\Gamma\left(\frac{D}{2}\right)}
\ v_{\text{i}}^{D-1} \ \Exp{-\frac{\alpha}{2} v_{\text{i}}^2}
                    \ \D v_{\text{i}} \ . \label{VeloDist}
\end{equation}
\section{The Ensemble Average} \label{SecAver}
\begin{figure} \begin{center}
\scalebox{0.55}{\epsffile{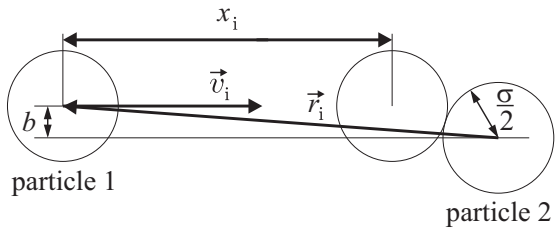}} \caption{The relative
coordinate system with respect to particle 2.} \label{ScatGen}
\end{center} \end{figure}

We determine the probability distribution for two particles
{under the condition that they will collide} in the future.
Therefore we depict the initial configuration of an arbitrary pair
of particles in relative coordinates
$\vec{r}_{\text{i}}=\vec{r}_1-\vec{r}_2$ as follows
(Fig.~\ref{ScatGen}): we rotate our coordinate frame such that the
horizontal axis is per definition
\begin{equation}
\vec{\text{e}}_x\equiv
\ \frac{\vec{v}_{\text{i}}}{v_{\text{i}}} \ , \label{Defex}
\end{equation}
with the initial relative velocity
$\vec{v}_{\text{i}}=\vec{v}_1-\vec{v}_2$.
This means that particle 2 rests in the origin while particle 1 
moves horizontally to the right.
Clearly, the particles will collide if and only if \label{Defb}
\begin{eqnarray*} 
(i)  && \text{the impact parameter is low enough, } \\
     && b=\sqrt{r_{\text{i}}^2-\left(\vec{r}_{\text{i}},\frac{\vec{v}_{\text{i}}}{v_{\text{i}}}\right)}\leq \sigma,  \\
(ii) && \text{and particle 1 is to the left of particle 2,
} \\
&& \left(\vec{r}_{\text{i}},\vec{v}_{\text{i}}\right)<0.
\end{eqnarray*}
For
{any} pair of velocities $\vec{v}_1, \vec{v}_2$, there
are initial relative spatial positions that lead to a collision.
So we have to integrate over the entire velocity space $\R^D\times
\R^D$,
\begin{equation}
\left(\frac{\alpha}{\pi}\right)^D
\int_{\R^D}\D^Dv_1\int_{\R^D}\D^Dv_2 \ \Exp{-\alpha
\left(v_1^2+v_2^2\right)}\ . \label{vInt}
\end{equation}
We take condition $(i)$ into account by integrating the impact
parameter over the interval $[0,\sigma]$. From the conventional
assumption of molecular chaos (i.e. the positions and velocities
of two particles are uncorrelated) follows that the impact is
uniformly distributed within the cross section,
\begin{equation}
P(b) \ \D b=(D-1) \frac{b^{D-2} \ \D b}{\sigma^{D-1}}, \
{0}<b<\sigma \ . \label{bInt}
\end{equation} Further, we need to know the horizontal
distance $x_{\text{i}}>0$ to the collision point. Together with
the impact parameter $b$ this determines the relative spatial
position
{completely} in the plane of incidence, since according to $(ii)$,
$\vec{r}=b \ \vec{\text{e}}_y-(x_{\text{i}}+\sqrt{\sigma^2-b^2}) \
\vec{\text{e}}_x$ always points to the left.

\begin{figure} \begin{center}
\scalebox{0.55}{\epsffile{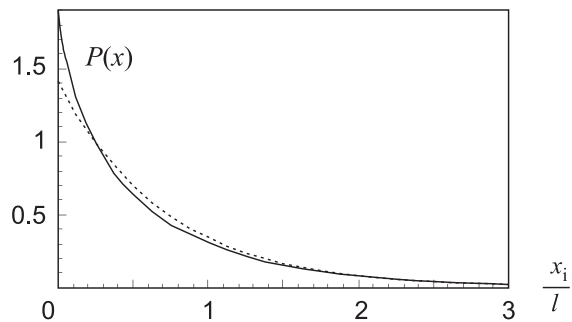}} \caption{The
distribution of $x_{\text{i}}$ after averaging out the velocities.
The dashed curve is an exponential distribution with the same
mean. Clearly $P(x_i)$ deviates from an exponential at distances
$x_{\text{i}}$ below the mean free path $l$.} \label{xDist}
\end{center} \end{figure}

The probability distribution of $x_{\text{i}}$ follows from the
distance covered by the particles in the laboratory frame.
Denoting by $x_1$ and $x_2$ the length that particle 1 and 2,
respectively, have travelled in the laboratory frame since the
beginning of the collision cycle, we have the equal time condition
\begin{equation}\frac{x_1}{v_1}=t_{\text{free}}=\frac{x_2}{v_2} \
,
\end{equation}
where $t_{\text{free}}$ stands for the time of free flight that
both particles have in common. From this follows for the initial
separation of particles
\begin{equation}
x_{\text{i}}=v_{\text{i}}\ t_{\text{free}} =
\frac{v_{\text{i}}}{v_1} x_1 \ .
\end{equation}
The probability density of the travelled distances $x_{1}$ and
$x_{2}$ are known in a gas to be
\begin{equation}
\Exp{-{x_j}/{l}} \ \frac{\D x_j}{l} \ , \quad j=1,2.
\end{equation}
The length scale $l$ is the mean free path in the
{laboratory frame}. Hence, under the assumption of molecular chaos
the probability density of the initial separation $x_{\text{i}}$
is
\begin{eqnarray*}
P(x_{\text{i}} \vert v_1,v_2) &=& C \int_0^\infty \frac{\D
x_{1}}{l} \ \int_0^\infty
\frac{\D x_{2}}{l} \ \Exp{-{(x_{1}+x_{2})}/{l}} \\
& & \qquad \times \
\delta\left(x_{\text{i}}-x_1\frac{v_{\text{i}}}{v_1}\right) \
\delta\left(\frac{x_1}{v_1}-\frac{x_2}{v_2}\right) \\
&=& C' \ \Exp{-\frac{x_{\text{i}}}{l} \
\frac{v_1+v_2}{v_{\text{i}}}}
\end{eqnarray*}
up to a normalization factor. Obviously this yields the
integration
\begin{equation}
\frac{v_1+v_2}{v_{\text{i}}} \int_0^\infty \frac{\D
x_{\text{i}}}{l} \ \Exp{-\frac{x_{\text{i}}}{l} \
\frac{v_1+v_2}{v_{\text{i}}}} \label{xInt}
\end{equation}
as part of the ensemble average. Putting (\ref{vInt}),
(\ref{bInt}) and (\ref{xInt}) together we can compute arbitrary
expectation values:
\begin{eqnarray}  \left < \dots \right>  &=&
(D-1) \left(\frac{\alpha}{\pi}\right)^D
\int_{\R^D}\D^Dv_1\int_{\R^D}\D^Dv_2 \
\frac{v_1+v_2}{v_{\text{i}}} \nonumber \\
&& \qquad \times \ \int_{0}^\sigma \frac{\D b \
b^{D-2}}{\sigma^{D-1}} \int_0^\infty \frac{\D
x_{\text{i}}}{l} \nonumber \\
&& \qquad \times \ \Exp{-\alpha \left(v_1^2+v_2^2\right)
-\frac{x_{\text{i}}}{l} \ \frac{v_1+v_2}{v_{\text{i}}} }
 \dots \label{average}
\end{eqnarray}
with $v_{\text{i}}=\Vert \vec{v}_1-\vec{v}_2\Vert$. In passing we
take a look at the 
distribution of $x_{\text{i}}$ in Fig.~\ref{xDist}. The joint
distribution (\ref{average}) implies that $x_{\text{i}}$ is
approximately distributed according to an exponential fall off, as
one may expect, because the distances in the laboratory frame
follow such a law. However there are differences: the mean is
lower, e.g. $\text{$<x_{\text{i}}>$} \approx 0.71 \ l$ for $D=2$,
and the distribution falls off faster than exponentially for small
$x_{\text{i}}$ (cf. \cite{Dorfman}).
\section{The Expansion of Velocity Space} \label{SecExpVelo}
We aim to compute the determinant of the matrix $M$ as defined by
Eq. (\ref{DefM}). There are always two distinct deviation matrices
$M_{\text{bound}}$ for $v_{\text{i}} < v_{\text{crit}}$ and
$M_{\text{scatt}}$ for $v_{\text{i}} > v_{\text{crit}}$, so that
the phase space average naturally decomposes into
\begin{eqnarray}
\left< \ln \vert \det M \vert \right >
 &=&\left< \ln \vert \det
M_{\text{bound}} \vert \right
>_{v_{\text{i}}<v_{\text{crit}}} \nonumber \\
&+& \left< \ln \vert \det M_{\text{scatt}} \vert \right
>_{v_{\text{i}} > v_{\text{crit}}} \ . \nonumber
\end{eqnarray}
After determining these matrices, Eq.~(\ref{KSe2}) will enable us
to compute
\begin{eqnarray}
 \frac{h_{\text{KS}}}{N}
 = \frac{\nu}{2} &{\big [}&\left< \ln
\vert \det M_{\text{bound}} \vert
\right>_{v_{\text{i}}<v_{\text{crit}}} \nonumber \\
 &+& \left< \ln \vert
\det M_{\text{scatt}} \vert \right >_{v_{\text{i}} >
v_{\text{crit}}} {\big ]} \ .
 \label{GenRes}
\end{eqnarray}
Because of momentum conservation,
$\vec{V}_{\text{i}}=\vec{V}_{\text{f}}$, the matrix $M$ is of the
blocked form
\begin{eqnarray*}
M=\left(\begin{array}{cc} \E_D & \Theta_D \\ \Theta_D & M'
\end{array}\right) \ , 
\end{eqnarray*}
where $\E_D$ and $\Theta_D$ are unity and zero matrices of
dimension $D\times D$ respectively. Therefore the only
contribution to the growth in velocity space stems from the
relative velocities,
\begin{equation}
\text{det}\; M = \text{det}\; M' \ . \label{mem_}
\end{equation}
The final relative velocity \footnote{Note that in this context
$v_{\text{loss}}$ is given by ${m}
v_{\text{loss}}^2/{4}=\phi+E_{\text{loss}}$ as a function of $r$
for the case of sticking particles, when 'final' does not refer to
the rupture event.} is
\begin{equation}
\vec{v}_{\text{f}}=\sqrt{v_{\text{i}}^2-v_{\text{loss}}^2}\left(\cos
\vartheta \ \e_x + \sin \vartheta \ \e_y \right) \ .
\label{FormulaFinal}
\end{equation}
As defined in (\ref{Defex}) $\e_x$ points in the direction of the
incoming velocity and $\e_y=\e_x \times \frac{\vec{r}\times
\vec{v}_{\text{i}}}{\Vert \vec{r}\times \vec{v}_{\text{i}}
\Vert}=\frac{\vec{r} v_{\text{i}}^2-\vec{v}_{\text{i}}\;
(\vec{r},\vec{v}_{\text{i}})}{\Vert \vec{r}
v_{\text{i}}^2-\vec{v}_{\text{i}}\; (\vec{r},\vec{v}_{\text{i}})
\Vert}$ is the orthogonal vector spanning the plan of motion, such
that
\begin{eqnarray*}
\vec{r}= -X_{\text{i}} \e_x + b \e_y
\end{eqnarray*}
with $X_{\text{i}}=x_{\text{i}}+x_{\text{col}}$ and
$x_{\text{col}}=-(\vec{r}_{\text{col}},\e_x)=\sqrt{\sigma^2-b^2}$
is the $x$-distance of the particles in the moment of collision.

When considering deviations of (\ref{FormulaFinal}) one has to
take into account contributions due to the change of the angle
\footnote{One has to distinguish between the variation of the
function $b$ (as given in item (i) on page \pageref{Defb}),
$\delta
b(\vec{r}_{\text{i}},\vec{v}_{\text{i}})=\delta(\vec{r}_{\text{i}},\e_y)=\delta
b - \frac{X_{\text{i}}}{v_{\text{i}}}\delta v_y$, and $\delta
b=(\delta \vec{r}_{\text{i}},\e_y)$ as an opportune notation for
the spatial deviation $\delta y$.}
$\vartheta=\vartheta(b(\vec{r},\vec{v}_{\text{i}}),v)$,
\begin{equation}
\delta \vartheta = \frac{\pd \vartheta}{\pd b} \delta b +
\frac{\pd \vartheta}{\pd b}
\frac{X_{\text{i}}}{v_{\text{i}}}\delta v_y+\frac{\pd
\vartheta}{\pd v_{\text{i}}} \delta v_x \ , \label{VarTheta}
\end{equation}
as well as contributions caused by rotations and inclinations of
the orbital plane of motion:
\begin{equation}
\left(\begin{array}{c} \delta \e_x \\  \delta \e_y \\  \delta  \e_z \\
\vdots
\end{array}\right) = \left( \begin{array}{cccc}
0 & \frac{\delta v_y}{v_{\text{i}}} & \frac{\delta
v_z}{v_{\text{i}}}
& \dots \\
- \frac{\delta v_y}{v_{\text{i}}} & 0 & \frac{X_{\text{i}}}{b} \frac{\delta v_z}{v_{\text{i}}} & \dots \\
- \frac{\delta v_z}{v_{\text{i}}} & -\frac{X_{\text{i}}}{b} \frac{\delta v_z}{v_{\text{i}}} & 0 \\
\vdots & \vdots & & \ddots
\end{array}\right)
\left(\begin{array}{c}\e_x \\ \e_y \\ \e_z \\ \vdots
\end{array}\right)
\ . \label{VarFrame}
\end{equation}
The Eqs. (\ref{VarTheta}) and (\ref{VarFrame}) hold for arbitrary
spatial dimension $D$. The resulting de\-via\-tion matrix $M'$ is
rather complicated:
\begin{widetext}
\begin{equation}
M'=\left( \begin{array}{ccccc} \frac{\cos
\vartheta}{\epsilon}-\epsilon v_{\text{i}} \vartheta_v \sin
\vartheta & -\left(1+X_{\text{i}} \vartheta_b\right) \epsilon \sin
\vartheta &
0 & \dots \\
\frac{\sin \vartheta}{\epsilon}+\epsilon v_{\text{i}} \vartheta_v
\cos \vartheta & +\left(1+X_{\text{i}} \vartheta_b \right)
\epsilon \cos \vartheta &
0 & \dots \\
0 & 0 & \epsilon \left( \cos \vartheta + \frac{X_{\text{i}}}{b} \sin \vartheta \right) & 0 & \dots \\
\vdots & \vdots & 0 & \epsilon \left( \cos \vartheta + \frac{X_{\text{i}}}{b} \sin \vartheta \right) \\
&&\vdots && \ddots
\end{array}\right)
\end{equation}
\end{widetext}
with the restitution coefficient (\ref{ResKoeff})
and the abbreviations $\vartheta_b\equiv\frac{\pd \vartheta}{\pd
b}$, $\vartheta_v\equiv\frac{\pd \vartheta}{\pd v}$. The
determinant of $M$ (which equals $M'$, cf. Eq.~(\ref{mem_})) is
surprisingly simple:
\begin{eqnarray} 
\text{det}\; M = \left(-1+x_{\text{i}}\frac{\pd \vartheta}{\pd
b}\right)\left(1-\frac{v^2_{\text{loss}}}{v^2_{\text{i}}}\right)^{\frac{D}{2}-1}
 \nonumber \\
 \times \left(1+\frac{x_{\text{i}}}{b}
\sin \vartheta \right)^{D-2}
 \ , \label{DetM_}
\end{eqnarray}
where we eliminated $x_{\text{coll}}\ll x_{\text{i}}$ using
\begin{eqnarray*} 
&& x_{\text{coll}} \vartheta_b \approx -2 \\
&& \frac{x_{\text{coll}}}{b} \sin \vartheta \approx 2-2\frac{b^2}{\sigma^2} \\
&& \cos \vartheta \approx 2\frac{b^2}{\sigma^2}-1 \ .
\end{eqnarray*}
This reduces in the dry case, $v_{\text{loss}}=0$, to the
expressions (18) ($D$=2) and (19) ($D$=3) in \cite{Dorfman}. The
first factor in (\ref{DetM_}) is always non-zero since $\frac{\pd
\vartheta}{\pd b}<0$.
\section{Results for the Kolmogorov-Sinai entropy} \label{SecRes}
\begin{figure} \begin{center}
\scalebox{0.25}{\epsffile{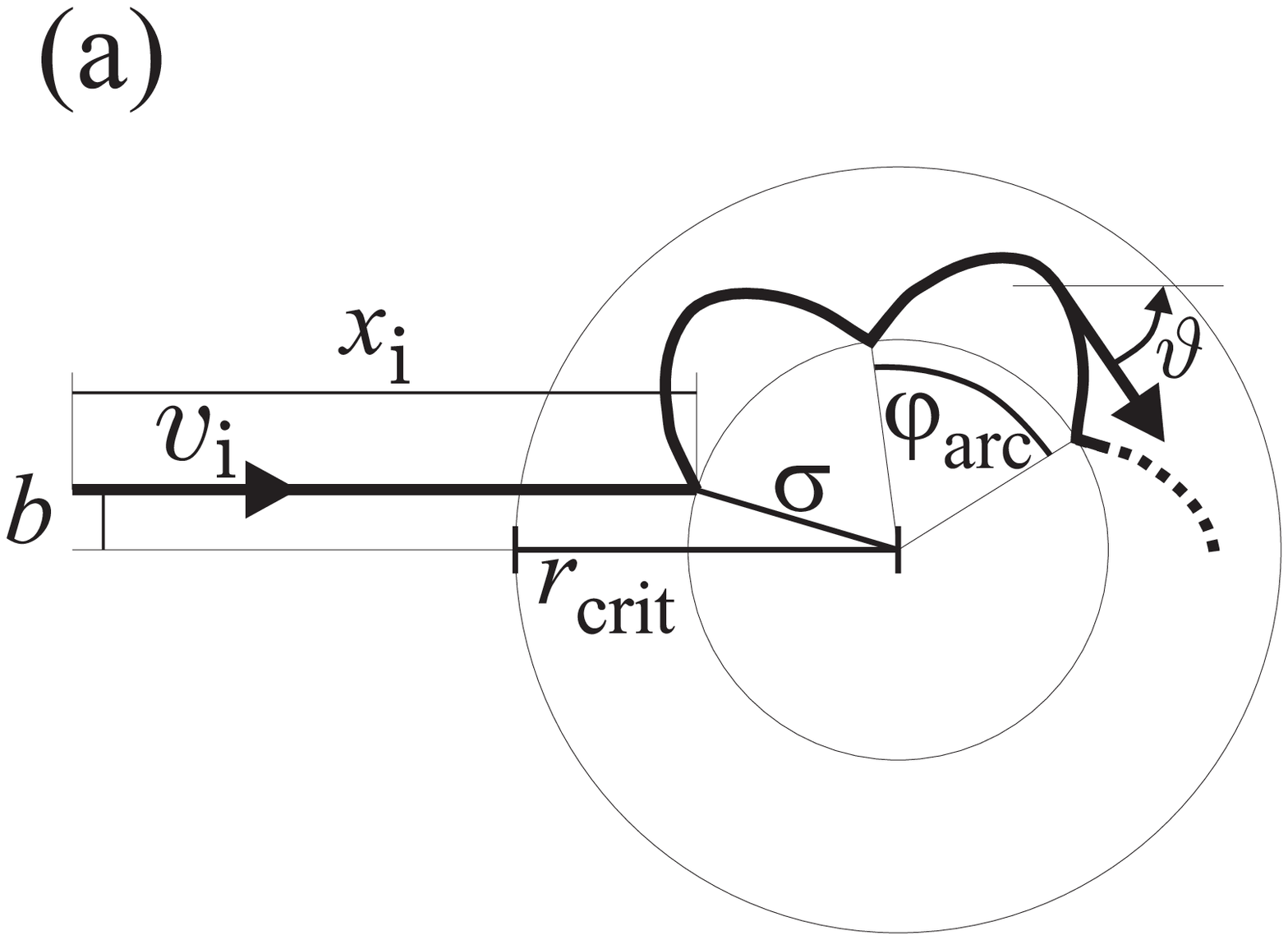}} \quad
\scalebox{0.25}{\epsffile{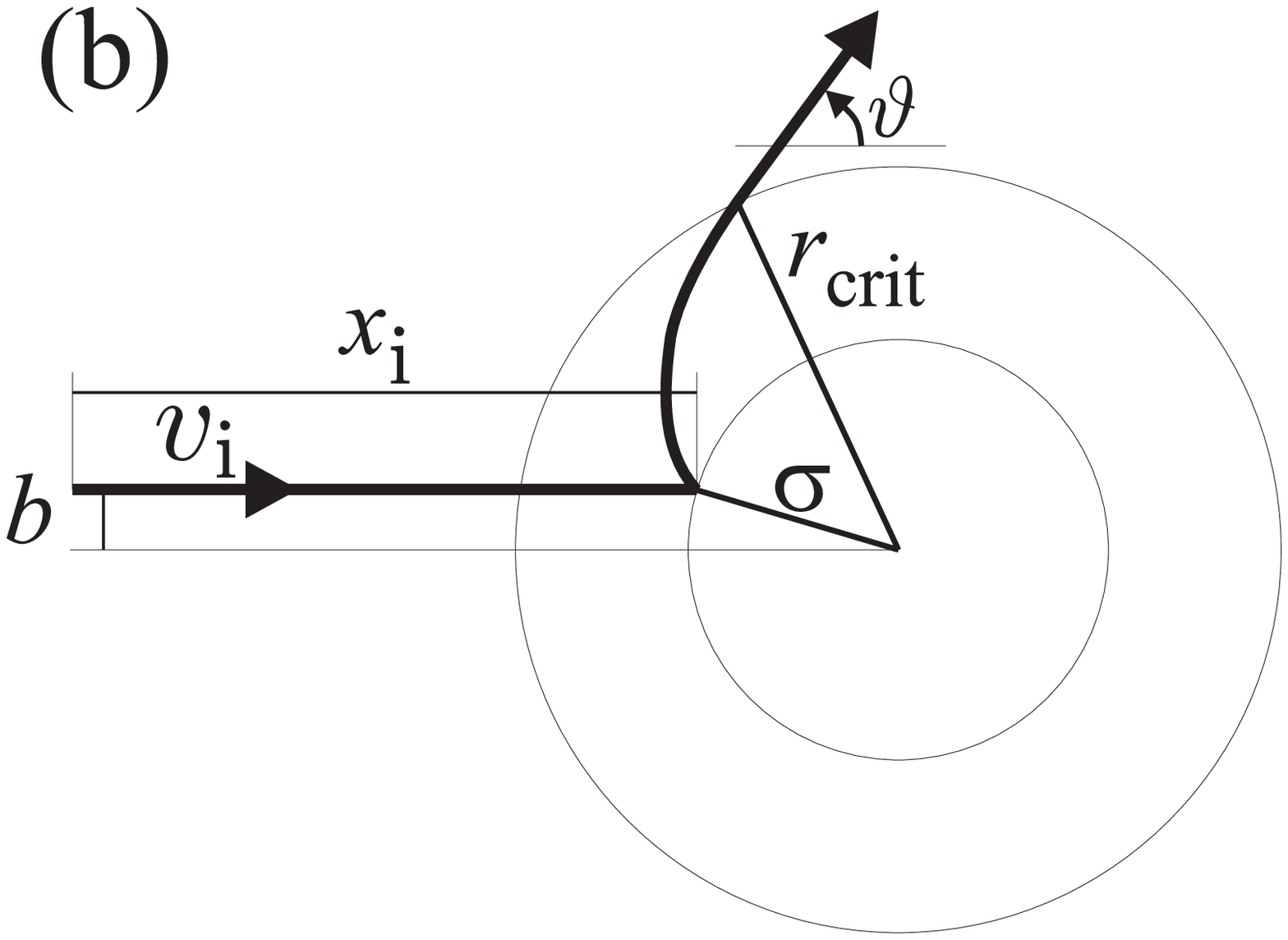}} \caption{The relative
motion for (a) sticking and (b) scattering.} \label{RelMot}
\end{center} \end{figure}

In Fig.~\ref{RelMot} the relative dynamic $\vec{r}(t)$ (which
equals the motion of one of the two particles in the center of
mass system up to a factor of 2) is sketched.
In both cases, the determinant of $M$ is of the form
(\ref{DetM_}),
but the meaning of the angle $\vartheta(b,v_{\text{i}})$ is quite
different. For impact velocities
{above} the critical value,
$\vartheta$ is the scattering angle
\begin{eqnarray} 
\vartheta_{\text{scatter}}(b,v_{\text{i}})
&=&\pi-\arcsin \frac{b}{\sigma} \nonumber \\
&-& \arcsin \frac{b}{r_{\text{crit}} \
\sqrt{1-\left(\frac{v_{\text{loss}}}{v_{\text{i}}}\right)^2}}
\nonumber \\
&-& \int_{\sigma}^{r_{\text{crit}}} \D\varphi_\phi(r)
 \ , \nonumber
\end{eqnarray}
whereas for $v_{\text{i}}<v_{\text{critical}}$ the angle
$\vartheta$ is a function of time,
\begin{eqnarray*}
\vartheta_{\text{bound}}(t_3,b,v_{\text{i}})&=&\frac{\pi}{2}-\arcsin\frac{b}{\sigma}
\\ &-& t_3\frac{\varphi_{\text{arc}}(b,v_{\text{i}})}{t_{\text{arc}}(b,v_{\text{i}})}
- \varphi_{\text{osc}}(t_3,b,v_{\text{i}}) \ .
\end{eqnarray*}
Here $t_3$ denotes the time during which the two-particle systems
remains bound until it is freed by a third particle. The angle
between two contacts $\varphi_{\text{arc}}(b,v_{\text{i}})$ equals
$2 \int_{\sigma}^{r_{\text{max}}(b,v_{\text{i}})}
\D\varphi_\phi(r)$ and there is a similar integral for the time
$t_{\text{arc}}$ it takes to run through one arc. The index $\phi$
ought to remind us that the potential (\ref{potential}) enters
only through these integral expressions. For $t_3\gg
t_{\text{arc}}$ the angle $\vartheta_{\text{bound}}$ grows
linearly with time, while the bound oscillations
$\varphi_{\text{osc}}$ are negligible.

Depending on the details of the interaction potential,
$\varphi_{\text{arc}}$ and $t_{\text{arc}}$ can grow beyond all
bounds as the pair $(b,v_{\text{i}})$ approaches the critical line
$(b,v_{\text{crit}}(b))$ (cf. Fig.~\ref{Gross}) in the bound
regime (from below). This singular behavior occurs in the Extended
Capillary Model (linear force, Fig.~\ref{LinForce_Fig}), whereas
in the Minimal Capillary Model (constant force) both quantities
remain finite. Close to the divergence the motion is an outward
directed spiral, so that the turning point is never reached and
the periodic collisions end. The interaction time can also diverge
for scattering states (reaching the critical line in
Fig.~\ref{Gross} from top), but this singularity is integrable
with respect to velocity. In the bound case the divergence is cut
off by the third particle and because of angular momentum
conservation we have the estimate
\begin{equation}
\vartheta_{\text{bound}}(t_3,b,v_{\text{i}}) \leq \text{const} +
\frac{t_3 b v_{\text{i}}}{\sigma^2}  \ . \label{ThBoundApprox}
\end{equation}
We will use the right-hand side as an approximation. The stopping
time $t_3$ is a random variable itself and distributed according
to
\begin{equation}
\frac{V_{\text{i}}}{l'} \ \Exp{-\frac{V_{\text{i}}}{l'}t_3} \ \D
t_3 \ , \label{t3dist}
\end{equation}
for a given center of mass velocity $V_{\text{i}}$ of the bound
system. There is a smaller mean free path $l'$ for the bound
two-particle system: since its total cross section changes with
time the effective diameter $\sigma_{\text{eff}}$ is
$\frac{3}{2}\sigma$ so that the mean center-center distance at
contact is $\frac{5}{4}\sigma$. Another factor of
$\sqrt{\frac{2}{3}}$ is caused by the mass ratio \cite{Gaspard},
thus
\begin{equation}
l' = \left(\frac{4}{5}\right)^{D-1} \sqrt{\frac{2}{3}} l \ .
\end{equation}
In the following, we shall evaluate averages that are linear in
$t_3$, so that we can forthwith substitute the expectation value,
$\overline{t_3}=\frac{l'}{V_{\text{i}}}$, of the distribution
(\ref{t3dist}).
Then from (\ref{ThBoundApprox}) follows
\begin{equation}
\frac{\pd \vartheta_{\text{bound}}}{\pd b}(v_{\text{i}}) \approx
\frac{v_{\text{i}} \ l'}{V_{\text{i}} \ \sigma^2}  \ . \label{611}
\end{equation}
In both cases, binding and scattering,
$\frac{\pd \vartheta}{\pd b}$ is at least of the order of $\frac{1}{\sigma}$%
, while $x_{\text{i}}$ is
of the order of the mean free path
\begin{equation} l = \frac{\Gamma\left(\frac{D+1}{2}\right)}{\sqrt{2} \pi^{\frac{D-1}{2}}} \left(\sigma^{D-1} n\right)^{-1}\
,
\end{equation}
with $n$ being the number density of grains.  Formulas for the
mean free path are well established \cite{Chernov} and other
charac\-teris\-tic quantities for the motion of tracer particles
are also available \cite{Gaspard}. We remark that investigating
the trajectories of tracer particles is a promising technique for
the experimental confirmation of results presented in this
article.

Our goal is to expand the KSE in the small dimensionless parameter
$n \sigma^D\ll 1$. So this is an expansion for the dilute wet
granular system. The unity in the first and the last factor in
(\ref {DetM_}) contributes to the KSE only in linear and higher
orders, while we are interested in the logarithmic and zeroth
order terms:
\begin{eqnarray}
\vert \det M \vert &=& x_{\text{i}} \ \left\vert \frac{\pd
\vartheta}{\pd b} \right\vert \
\left(1-\frac{v^2_{\text{loss}}}{v^2_{\text{i}}}{\rm
\theta}(v_{\text{i}}-v_{\text{crit}})\right)^{\frac{D}{2}-1} \nonumber \\
&& \times \left(\frac{x_{\text{i}}}{b} \sin \vartheta
\right)^{D-2} . \label{DetM}
\end{eqnarray}
With the step function $\theta$, Eq.~(\ref{DetM}) is valid for
scattering and binding, because we assume that the collision with
the third particle rethermalize the two-particle system, so that
the next collision cycle starts with the same initial
distribution. Since the 'third' particles have an energy of the
order of the granular temperature $T \gg E_{\text{loss}}$ we can
safely neglect the formation of bound states of three or more
particles (cf. Fig.~\ref{CoordinationPlot}). A cluster size
expansion will be discussed at the end of this section.

After introducing the appropriate length scales $l$ and $\sigma$
we are lead to examine
\begin{eqnarray} \label{2DKSE}
\frac{h_{\text{KS}}}{N} &=& \frac{\nu}{2} {\Big [} (D-1) \ln
\frac{l}{\sigma} - (D-2) \left< \ln \frac{b}{\sigma} \right> \nonumber \\
 &+& (D-1) \left< \ln \frac{x_{\text{i}}}{l}
\right> \nonumber \\
&+& \left<
\ln \left( \sigma \ \left\vert \frac{\pd
\vartheta_{\text{bound}}}{\pd b} \right\vert \right)
\right>_{v_{\text{i}} < v_{\text{crit}}} \nonumber \\
 & + & \left<{\left(\frac{D}{2}-1\right)} \ln \epsilon +
\ln \left( \sigma \ \left\vert \frac{\pd
\vartheta_{\text{scatt}}}{\pd b} \right\vert \right) \right
>_{v_{\text{i}} > v_{\text{crit}}} \nonumber \\
& + & (D-2) \left< \ln \vert \sin \vartheta \vert \right>  {\Big
]} \ .
\end{eqnarray}
The 
{first two terms} in the square bracket yield
\begin{equation}
- \ln n\sigma^D-C_D \ , \label{LT1}
\end{equation}
with a numerical constant $C_D=\frac{D-1}{2}\ln
2+\frac{(D-1)^2}{2}\ln \pi - \frac{D-2}{D-1}-(D-1)\ln
\Gamma\left(\frac{D+1}{2}\right)$. This is independent of the
ensemble average and the interaction potential.

If $x_{\text{i}}$ was distributed exponentially with
mean $l$, the 
{third term} in (\ref{2DKSE}) would give rise to the negative of
Euler's constant, $- \gamma_{\text{Euler}} \approx -0.5772$,
independent of the dimensionality of the problem. As discussed
before, lower values of $x_{\text{i}}$ are favored. That is why we
find by numerical computation a lower expectation value, e.g. for
$D=2$:
\begin{equation}
\left< \ln \frac{x_{\text{i}}}{l} \right> \approx -1.01.
\end{equation}
The 
{fourth term} in (\ref{2DKSE}) is (cf. Eq.~(\ref{611}))
\begin{eqnarray}
&&\left<
\ln \left( \sigma \ \left\vert \frac{\pd
\vartheta_{\text{bound}}}{\pd b} \right\vert \right)
\right>_{v_{\text{i}} < v_{\text{crit}}} \nonumber \\ &=& -
\left(\ln n \sigma^D+\tilde C_D \right)
\left<1\right>_{v_{\text{i}} < v_{\text{crit}}} \nonumber
\\
&& + \left< \ln \frac{v_{\text{i}}}{V_{\text{i}}}
\right>_{v_{\text{i}} < v_{\text{crit}}} \ , \label{LT2}
\end{eqnarray}
with the numerical constant $\tilde C_D=(D-1)\ln
\frac{5}{4}+\frac{\ln 3}{2}+\frac{D-1}{2}\ln \pi -\ln
\Gamma\left(\frac{D+1}{2}\right)$.

Together with (\ref{LT1}) the logarithm $\ln n\sigma^D$ herein
forms the leading term of the density expansion. Therefore the
logarithm $\ln n\sigma^D$ in (\ref{LT2}) is a correction of the
leading term as it is known for the dry case \cite{Dorfman}. The
KSE has the following density expansion:
\begin{eqnarray}
\frac{h_{\text{KS}}}{N}&=&-{\nu} A_D  \ln n \sigma^D + \nu B_D +
\mathcal{O}(n\sigma^D),  \label{Expansion}
\end{eqnarray}
with the leading coefficient
\begin{eqnarray}
A_D&=&A_D\left(
\frac{E_{\text{loss}}}{T},\frac{r_{\text{crit}}}{\sigma}\right)
\nonumber \\ &=&
\frac{D-1}{2}+\frac{D-1}{\Gamma(\frac{D}{2})}\left(\frac{m}{4T}\right)^{\frac{D}{2}}
\nonumber \\ &\times& \int_0^\sigma \frac{\text{d} b \
b^{D-2}}{\sigma^{D-1}} \ \int_0^{v_{\text{crit}}(b)} \text{d} v \
v^{D-1} \ \Exp{-\frac{m}{4T}v^2} \ ,
\end{eqnarray}
and the density independent part
\begin{eqnarray}
B_D& =\frac{1}{2}\big{[}- C_D + (D-1)
\left<\ln\frac{x_{\text{i}}}{l}\right> - \tilde C_D
\left<1\right>_{v_{\text{i}} < v_{\text{crit}}} \nonumber \\ & +
\left< \ln \frac{v_{\text{i}}}{V_{\text{i}}} \right>_{v_{\text{i}}
< v_{\text{crit}}} + \left< \ln \left( \sigma \ \left\vert
\frac{\partial \vartheta_{\text{scatt}}}{\partial b} \right\vert
\right) \right>_{v_{\text{i}}
> v_{\text{crit}}}
\nonumber \\& + ({D}-2) \left(\frac{\left< \ln \epsilon
\right>_{v_{\text{i}}>v_{\text{crit}}} }{2} + \left< \ln \vert
\sin \vartheta \vert \right>  \right) \big{]}. \label{BCoeffForm}
\end{eqnarray}
\begin{figure} \begin{center}
\scalebox{0.7}{\epsffile{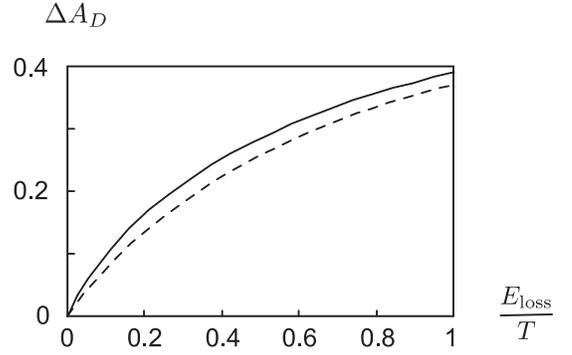}} \caption{The increase
$\Delta A_D=\frac{P_{\text{bound}}}{2}$ of the leading coefficient
$A_D=\frac{D-1}{2}+\Delta A_D$:
The solid line is for two, the dashed line for three dimensions
$D$.
Since $A=\frac{D-1}{2}$ in the absence of the liquid bridge interaction 
we recover the result for dry granulates as a special case. With
the approximation for the wet granular gas used in the derivations
one is restricted to temperatures above the bridge energy
$E_{\text{loss}}$. Otherwise the method applied has to be extended
to take clusters of more than two particles sticking together into
account. The far extreme case, $E_{\text{loss}}\gg T$, is known as
the so-called sticky gas.
} \label{LeadingT} 
\end{center} \end{figure}

The general form of the leading term, valid for any velocity
distribution, is
\begin{eqnarray}
A_D=\frac{D-1}{2}+\frac{P_{\text{bound}}}{2} \ .
\label{GeneralResult}
\end{eqnarray}

We want to emphasize that so far all results of this section are
general with respect to the spatial dimensionality of the problem
and the details of the particle interaction. The probability
$P_{\text{bound}}=\left<1\right>_{v_{\text{i}} < v_{\text{crit}}}$
in (\ref{GeneralResult}) is given by integrating velocity and
impact factor over the bound states in Fig.~\ref{Gross}. Only here
the detailed interaction models (\ref{ECMkrit}) and
(\ref{MCMkrit}) enter the problem.

Let us now turn to explicit results. For the Gaussian velocity
distribution (\ref{VeloDist})
and odd spatial dimensions the velocity integral of
$P_{\text{bound}}$ is an incomplete Gamma function. In even
dimensions the integral is elementary, yielding for $D=2$
\begin{eqnarray*}
A_2(\varepsilon,\gamma)&=1-\frac{1}{2} \int_0^1 \D x \
\Exp{-\varepsilon \ f(x,\gamma)} , \ \
\varepsilon=\frac{E_{\text{loss}}}{T} ,
\end{eqnarray*}
as a function of the bridge energy over granular temperature,
$\varepsilon$, and the wetting content,
$\gamma={r_{\text{crit}}}/{\sigma}\geq 1$. The remaining
integration variable is the impact parameter, $x={b}/{\sigma}$.
The excess of the critical energy over the bridge energy,
$f(x,\gamma)={E_{\text{crit}}}/{E_{\text{loss}}}$, depends on the
model details. In the Minimal Capillary Model from
Eq.~(\ref{MCMkrit}) follows
\begin{eqnarray*}
f(x,\gamma)=\left({1-\frac{x^2}{\gamma^2}}\right)^{-1} \ .
\end{eqnarray*}
The coefficient $A_D$ of the Minimal Capillary Model is plotted in
Fig.~\ref{LeadingT} as a function of the liquid bridge energy for
two and three dimensions. Very similar curves follow from the
Extended Capillary Model. For the plot the limit of short liquid
bridges, ${r_{\text{crit}}}={\sigma}$, was chosen. This
corresponds to a small amount of liquid that is just sufficient to
wet the surface roughness of realistic spheres. Independent of
$r_{\text{crit}}/\sigma \geq 1$, in the dry limit (or equivalently
the high temperature limit) $A_D$ approaches $(D-1)/2$, which is
the known result for hard spheres \cite{Dorfman}. For a higher
content of wetting liquid, $r_{\text{crit}}/\sigma > 1$, the
dependence of the leading term on the binding energy becomes
flatter, but in an experimental situation there is a simultaneous
gain in $E_{\text{loss}}$ when liquid is added. Varying the
surface tension of water by adding a salt to the wetting solution
is an experimentally feasible way to measure this curve directly
with a fixed amount of wetting liquid, such that
$r_{\text{crit}}/\sigma$ can be kept constant.

From this graph we see the sensitive dependence of the KSE on the
cohesion force of the wetting liquid. To gain analytic insight we
investigate exemplarily the two-dimensional case plotted.
Substituting $z=1/(1-x^2)$ gives
\begin{eqnarray}
A_2(\varepsilon,1)=1-\frac{1}{4}\int_1^\infty \frac{\D z}{z^2}
\frac{\Exp{-\varepsilon z}}{\sqrt{1-1/z}}
\end{eqnarray}
Splitting up the integration at $z=1/\varepsilon$ allows to
separate the non-analytic part.
\begin{eqnarray}
A_2(\varepsilon,1)=1 &-&\frac{\varepsilon}{4}\int_1^\infty
\frac{\D z}{z^2}
\frac{\Exp{-z}}{\sqrt{1-\varepsilon/z}} \nonumber \\
 &-&\frac{1}{4}\int_1^\frac{1}{\varepsilon} \frac{\D z}{z^2}
\frac{\Exp{-\varepsilon z}}{\sqrt{1-1/z}} \label{ZerlInt}
\end{eqnarray}
The first integral in (\ref{ZerlInt}) can be expanded in powers of
$\varepsilon\in[0,1)$ since $z>1$. The second integral equals $2$
for $\varepsilon\rightarrow 0$, while its first derivative has a
logarithmic divergence:
\begin{eqnarray}
A_2(\varepsilon,1)=\frac{1}{2} + \varepsilon \left(C-\frac{\ln
\varepsilon}{4}\right)+{\cal O}(\varepsilon^2) \ .
\end{eqnarray}
The constant $C$ is $\int_1^\infty \exp{(-z)}/4z + {\ln 2}/{2} +
(1-1/\Exp{})/4 \approx 0.56$. This shows that the slope of $A_2$
is vertical at $E_{\text{loss}}=0$.

\begin{figure} \begin{center}
\scalebox{0.7}{\epsffile{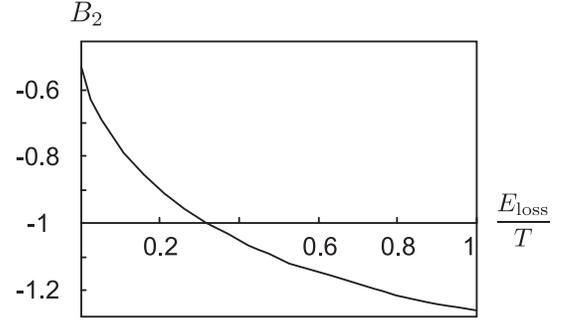}} \caption{The coefficient
$B_2$ of the density expansion (\ref{Expansion}).} \label{BCoeff}
\end{center} \end{figure}

Let us finally look at the next higher order term $B_D$ of the
density expansion. For simplicity we restrict ourselves to the
case $D=2$, so that
\begin{eqnarray}
B_2 &=& \frac{1}{2}\left[- C_2 +
\left<\ln\frac{x_{\text{i}}}{l}\right> - \tilde C_2
\left<1\right>_{v_{\text{i}} < v_{\text{crit}}} + \left< \ln
\frac{v_{\text{i}}}{V_{\text{i}}} \right>_{v_{\text{i}} <
v_{\text{crit}}} \right. \nonumber \\ &+& \left. \left< \ln \left(
\sigma \ \left\vert \frac{\partial
\vartheta_{\text{scatt}}}{\partial b} \right\vert \right)
\right>_{v_{\text{i}}
> v_{\text{crit}}} \ \right ]. \label{B2}
\end{eqnarray}
The {last term} in (\ref{B2}) is exactly equal to unity in the
limit of dry granulates,
\begin{eqnarray*}
 \lim_{E_{\text{loss}}\rightarrow 0}  \left< \ln \left( \sigma \
\left\vert \frac{\pd \vartheta_{\text{scatt}}}{\pd b} \right\vert
\right) \right
>_{v_{\text{i}} > v_{\text{crit}}} \\
= \int_0^\sigma \frac{\D b}{\sigma} \ \ln
\frac{2}{\sqrt{1-\left(\frac{b}{\sigma}\right)^2}}  = 1 \ ,
\end{eqnarray*}
but decreases as the critical velocity increases when we turn on
the liquid bridge interaction. The coefficient $B_2$ for the
zeroth order in the expansion (\ref{Expansion}) is plotted in
Fig.~\ref{BCoeff}.
It is known for the dry limit \cite{Dorfman}, that the accordance
of $B_D$ with numerical simulation cannot keep up with the
successful confirmation of $A_D$. The origin of this discrepancy
is the assumption that the unstable manifold coincides with
velocity space and it is quite involved to improve on that
\cite{Astrid}. In the dry limit our method yields $B_2=-0.52(8)$,
which is lower than the analytical estimate ($B_2=0.1045$) and the
simulated result 
($B_2=0.679$) of \cite{Dorfman}.

\begin{figure} \begin{center}
\scalebox{0.52}{\epsffile{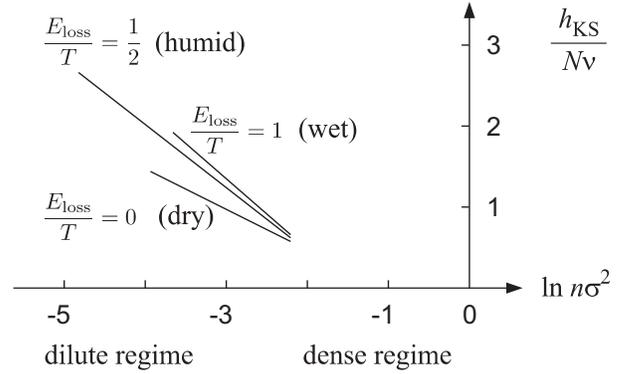}} \caption{The
two-dimensional KSE as a function of the density for three
different bridge energies $E_{\text{loss}}$. This energy depends
on the amount of wetting liquid added to the granular gas as is
indicated in the plot. Another way to change $E_{\text{loss}}$ is
to add a salt or a surfactant.} \label{VasiPlot}
\end{center} \end{figure}

From the knowledge of the coefficients $A_D$ and $B_D$ follows the
KSE in the dilute system for various wetting contents as shown in
Fig.~\ref{VasiPlot} for $D=2$.

\subsection*{The Cluster Expansion}
\begin{figure} \begin{center}
\scalebox{0.90}{\epsffile{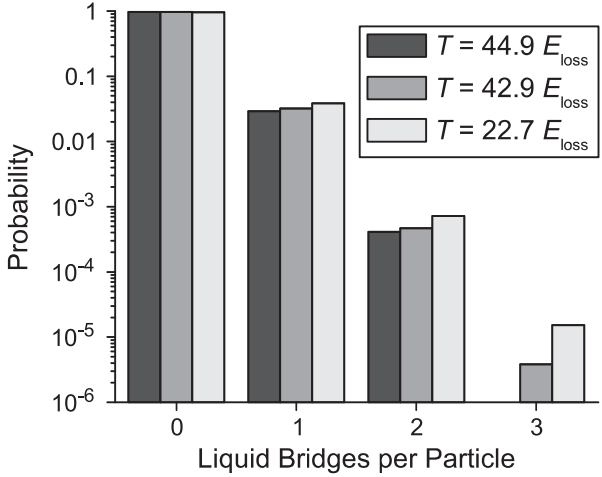}} \caption{The
probability for a sphere to have a certain number of liquid bonds
ending on its surface. This distribution is measured in a three
dimensional molecular dynamics simulation of a wet granular gas
with an occupied volume fraction of 3.9\%, which corresponds to $n
\sigma^3=0.074$. The granular temperature $T$ has been varied as
indicated. The probability for two liquid bridges ending on one
particle, as necessary for a three-particle-cluster, is suppressed
by more than three orders of magnitude. An analytic approach to
the KSE is favorable because the direct numerical integration
suffers from high computing times for the full tangent space
dynamics and yields noisy results \cite{LyapFluc,footnote}. The
liquid bond distribution shown is a robust and reliable
single-particle quantity.} \label{CoordinationPlot}
\end{center} \end{figure}
In Eq.~(\ref{GenRes}) we considered events including bound states
of two particles ($\text{a + b + c} \rightarrow \text{ab + c}
\rightarrow \text{a + b + c}$) and scattering events ($\text{a +
b}\rightarrow \text{a + b}$) by writing
\begin{eqnarray}
\left< \ln \vert \det M \vert \right> &=& \left< \ln \vert \det
M_{\text{bound}} \vert \right>_{v_{\text{i}} < v_{\text{crit}}}
\nonumber \\ &&+ \left< \ln \vert \det M_{\text{scatt}} \vert
\right>_{v_{\text{i}}
> v_{\text{crit}}} \ . \label{ShortScheme}
\end{eqnarray}
The first term is proportional to $P_{\text{bound}}$ which led to
Eq.~(\ref{GeneralResult}). Here we wish to point out how to
generalize the computation of the KSE to include clusters of
higher particle number. All equalities in~(10) hold for arbitrary
types of events, when $M_i$ denotes the deviation matrix
associated with the $i$th event and $\nu$ is the generalized event
frequency. Referring to the event type by $\text{T}$ we reorder
the averaging. Collecting the events of type T by introducing
$\delta_{\text{type}(j),\text{T}}$ (which is unity for an event T
and otherwise zero) we write $\left<\dots\right>_{\text{T}}$ for
$\left<\dots \ \delta_{\text{type}(j),\text{T}}\right>$:

\begin{equation}
\frac{2}{\nu N} \ h_{\text{KS}} = \left< \ln \vert \det M \vert
\right>  = \sum_{\text{T}} \ \left< \ln \vert \det
M_{\text{T}}\vert \right>_{\text{T}} \label{GeneralScheme} \ .
\end{equation}
The summation can be written as a systematic expansion in the
cluster size:
\begin{eqnarray*}
\left\{
\begin{array}{lllc}
\text{a + b} \rightarrow \text{a + b} &&& (\text{T}_1)\\
& & \text{a + b + c}  & (\text{T}_2)\\
&                    \quad  \nearrow \\
  \text{a + b + c} \rightarrow \text{ab + c} & \quad \rightarrow &
         \left\{\begin{array}{lr}
          \text{ac + b}\\
          \text{bc + a}\\
          \text{ab + c}\\
          \end{array} \right.  & \begin{array}{c} (\text{T}_3) \\ (\text{T}_4) \\ (\text{T}_5) \end{array}  \\
& \quad \searrow \\
& &  \text{abc} & (\text{T}_6)\\
\vdots
\end{array} \right.
\end{eqnarray*}
with the events $\text{T}_1$ and $\text{T}_2$ considered before in
(\ref{ShortScheme}). The events $\text{T}_j$ with $j>2$ result in
new many-particle-clusters which are exponentially rare components
of the wet granular gas as is evident from
Fig.~\ref{CoordinationPlot}. We remark that the scattering of a
bound state ($\text{T}_5$) prolongs the mean bond time $t_3$ to
become $t'_3=\alpha t_3$, with $\alpha=1 + 2 P_{\text{T}_5} + 3
P^2_{\text{T}_5} + \dots = 1/(1-P_{\text{T}_5})^2$. The unity in
front of this series corresponds to breaking the bound state in
its first collision ($\text{T}_2$), the second term corresponds to
one scattering event of the bound pair and the following terms to
multiscattering. The contribution to the KSE is proportional to
the logarithm of this time, $\ln t'_3 = \ln t_3 - 2 \ln
(1-P_{\text{T}_5})$. The first term $\ln t_3 \propto - \ln (n
\sigma^D)$ is the wet granular contribution to the leading
coefficient $A$ as identified in Eq.~(44). The second term gives a
correction to the $B$-coefficient which is of the order
$P_{\text{T}_5} = {\mathcal
{O}}\left(\sqrt{E_{\text{loss}}/T}^3\right)$ for three dimensions.

\section{Conclusions}
\subsection*{Summary}
We worked out the crucial difference in the interaction of wet
granulates compared to the dry case. There is a liquid bridge
causing a radial hysteretic force over finite distance. The
detailed distance dependence is of minor importance. The decisive
ingredient in the Capillary Model is the extraction of a bridge
energy that is independent of the initial velocity in contrast to
the ``Standard Model'' using a restitution to extract a certain
fraction of energy.

We found an enhanced chaotic behavior of the wet granular system.
The leading term in the expansion of the KSE with respect to the
small density ($n \sigma^2 \ll 1$) changed due to the possible
sticking of particles. One can think of the prolonged interaction
time enforcing the exponential separation in velocity space. The
continuous but in general not differentiable transition to the
limiting dry case has been established.

This dynamical property recommends the wet granular system as a
suitable candidate for experimental, numerical and analytic tests
of the Gallavotti-Cohen fluctuation theorem \cite{GCFT} which
requires hard chaos.

\subsection*{Outlook}
In this analytic work we used an assumption on the unstable
manifold and we neglected correlation effects in consecutive
collisions. Although physically motivated, the next challenge will
be to verify these assumptions by direct numerical simulations.

The rigorous derivation of phenomenological laws such as the
Navier-Stokes equation for viscous flow and the Fourier law for
heat transport is a fundamental problem under intense discussion.
Relations between the Lyapunov spectrum of the microscopic
dynamics and macroscopic properties such as viscosity and heat
conductivity have been established within the last years, most
detailed for the Lorentz gas
\cite{TC1,TC2a,TC2b,TC2c,TC2d,TC2e,TC3}. The severity and
importance of these relations become apparent from the fact they
have to bridge the gap between microscopic reversibility and
macroscopic irreversibility challenging physicists since Ludwig
Boltzmann.

The dynamics of the wet granular system studied in this work
follows a mesoscopic law including dissipation, and kinetic theory
has already been extended to dry granular matter \cite{Poeschel}.
The next step is to extend also these transport relations. We hope
that our results on the Lyapunov exponents might stimulate this
development. On the experimental side mechanical properties of wet
granulates are presently under investigation \cite{Mario}.

A further interesting problem is the computation of the KSE for
dense wet granulates. This might lead to a novel description of
clustering -- as a non-equilibrium phase transition -- in terms of
the Lyapunov spectrum. Yet this problem is challenging as it needs
new concepts, because the identification of the velocity space
with the instable manifold is limited to the dilute gas.

\begin{acknowledgments}
We thank H.~van~Beijeren and H.~Schanz for fruitful discussions.
A.~F. gratefully acknowledges the interaction with K.~R\"oller on
the simulation results shown in Fig.~\ref{CoordinationPlot}.
\end{acknowledgments}


\begin{thebibliography}{van Beijeren et al.}
\bibitem{footnote}
The direct integration algorithm has to adapt the time steps to
allow for an accurate integration of the extremely short contact
repulsion. With these fine integration steps it takes extremely
long until the system has explored its high dimensional phase
space as necessary for reliable Lyapunov spectra. Associated with
the high number of integration steps is an increase of the
numerical error. We remark that the superior event-driven
technique is not applicable for the Capillary Model of wet
granular matter in $D \geq 2$ dimensions.
\end{thebibliography}
\end{document}